%% file: main.tex
\pdfoutput=1 
\documentclass{trbunofficial}  

\usepackage{xcolor}
\usepackage{graphicx}
\usepackage{amsmath,amsfonts,amssymb,amsthm,epsfig,epstopdf,url,array}
\usepackage{amsfonts}
\usepackage{float}
\usepackage{subfig}
\usepackage{ragged2e}
\theoremstyle{plain}

\newtheorem{thm}{Theorem}[section]

\newtheorem{prop}[thm]{Proposition}

\theoremstyle{definition}

\theoremstyle{remark}

\AuthorHeaders{Qian, Xue, Ukkusuri}
\title{Modeling disease spreading with adaptive behavior considering local and global information dissemination}
\author{%
  \textbf{Xinwu Qian, Ph.D.}\\
  Lyles School of Civil Engineering, Purdue University\\
  550 Stadium Mall Dr, West Lafayette, IN, 47907\\
  qian39@purdue.edu\\
  \hfill\break
  \textbf{Jiawei Xue}\\
  Lyles School of Civil Engineering, Purdue University\\
  550 Stadium Mall Dr, West Lafayette, IN, 47907\\
  xue120@purdue.edu\\
  \hfill\break
    \textbf{Satish V. Ukkusuri, Ph.D.}\\
  Lyles School of Civil Engineering, Purdue University\\
  550 Stadium Mall Dr, West Lafayette, IN, 47907\\
  sukkusur@purdue.edu\\
(Corresponding Author)
}

\begin{document}

\begingroup
\def\uppercasenonmath#1{} 
\let\MakeUppercase\relax 
\maketitle
\endgroup

\section{Abstract}
The study proposes a modeling framework for investigating the disease dynamics with adaptive human behavior during a disease outbreak, considering the impacts of both local observations and global information. One important application scenario is that commuters may adjust their behavior upon observing the symptoms and countermeasures from their physical contacts during travel, thus altering the trajectories of a disease outbreak. We introduce the heterogeneous mean-field (HMF) approach in a multiplex network setting to jointly model the spreading dynamics of the infectious disease in the contact network and the dissemination dynamics of information in the observation network. The disease spreading is captured using the classic susceptible-infectious-susceptible (SIS) process, while an SIS-alike process models the spread of awareness termed as unaware-aware-unaware (UAU). And the use of multiplex network helps capture the interplay between disease spreading and information dissemination, and how the dynamics of one may affect the other. Theoretical analyses suggest that there are three potential equilibrium states, depending on the percolation strength of diseases and information. The dissemination of information may help shape herd immunity among the population, thus suppressing and eradicating the disease outbreak. Finally, numerical experiments using the contact networks among metro travelers are provided to shed light on the disease and information dynamics in the real-world scenarios and gain insights on the resilience of transportation system against the risk of infectious diseases. 
\hfill\break%
\noindent\textit{Keywords}: Disease spreading, information dissemination, multiplex network, adaptive behavior, transportation contact network
\newpage

\input{introduction.tex}
\input{notation.tex}
\input{assumption.tex}
\input{ua_sis.tex}
\input{formulation.tex}
\input{analysis.tex}

\input{results.tex}
\input{conclusion.tex}

\section{Author Contributions}
The authors confirm contribution to the paper as follows: study conception and design: X. Qian, S.V. Ukkusuri; data collection: X.Qian; analysis and interpretation of results: X. Qian, J. Xue; draft manuscript preparation: X. Qian, J. Xue, S.V. Ukkusuri. All authors reviewed the results and approved the final version of the manuscript.

\bibliographystyle{unsrt}
\bibliography{ref}
\end{document}

%% file: introduction.tex
\section{Introduction}

People are now engaged in more intensive daily activities and exposed to massive information from various sources than ever before. 
However, one negative consequence of intensive activities is the accelerated outbreaks of infectious diseases through frequent travels and the exchange of goods, which turns small-scale local transmissions in the past into large-scale global pandemics such as the SARS, H1N1, MERS and more recently the COVID-19. On the other hand, the exchange of vast local and global information offers the public a distinct opportunity to track and monitor the state of ongoing disease outbreaks. One particular example is the global outbreak of the COVID-19, where we have witnessed the number of infections growing at an unprecedented pace, and there are now over 16.3 million confirmed infections worldwide\cite{2020who,2020cases}. Meanwhile, numerous COVID-19 monitoring dashboards are gaining popularity among the public~\cite{dong2020interactive} and the epidemic has been the focus of the state media and social media platforms. As intensive activities are deemed to facilitate the spread of infectious diseases with frequent encounters, the exposure to mass information, however, may contribute to reshaping the activity patterns and how travelers are getting into contact and eventually altering the fate of disease outbreaks. This motivates us to investigate the disease dynamics in the contact network with adaptive travelers' behavior, considering the co-evolution of disease and information dynamics. 


Numerous studies investigated disease percolation over networks with individuals as the nodes and the connections between them (e.g., physical encounters) as the edges, and detailed reviews of related works can be found in~\cite{meyers2007contact,danon2011networks}. One important threshold value for network epidemiology is the disease threshold $\lambda_{c}$, whose value determines the stable state of a disease outbreak. Studies have shown that $\lambda_{c}$ is a function of the degree distribution of the network~\cite{pastor2001epidemic,moreno2002epidemic} and is inversely proportional to the eigenvalue of the adjacency matrix of the network~\cite{youssef2011individual,granell2014competing}, which motivates subsequent studies to investigate disease dynamics under different network layouts. Bogua et al.~\cite{bogua2003epidemic} modeled spreading diseases using susceptible-infectious-susceptible (SIS) model in uncorrelated and correlated networks and established a concise analytical expression for $\lambda_{c}$ in the uncorrelated network and used the eigenvalue of the network matrix to describe $\lambda_{c}$ when network is correlated. 
Dickison et al.~\cite{dickison2012epidemics} divided individuals into two inter-connected network layers and focused on disease spreading within and between the two layers using simulations. Sanz et al.~\cite{sanz2014dynamics} separated the dynamics of two diseases into two networks, investigated the correlated states between the two network layers, and established the analytical expression for the disease threshold. In addition, disease spreading in two networks with partially overlapped nodes was researched by Buono et al.~\cite{buono2014epidemics}. 

In general, the co-evolution of disease dynamics and information dynamics may be investigated following the same inter-correlated multiplex network setting, as in the above-mentioned studies. Several recent studies made the initial attempts to address this issue. For instance, Wang et al. modeled information and disease outbreak in communication and contact layers, respectively. They assumed that one would be vaccinated if and only if they are aware of the disease~\cite{wang2014asymmetrically}, and they later extended their work by introducing the analytical threshold for disease-infected neighborhoods and improving the original vaccination strategy~\cite{wang2016suppressing}. Kabir et al~\cite{kabir2019analysis} classified the whole population into six categories based on two awareness states (unaware, aware) and three infection states (susceptible, infected, recovered), and they introduced eight parameters to model the state transition among the six categories. Similarly,  Xia et al. modeled the state transitions between five states of awareness and disease infection \cite{xia2019new}. They assumed that the infected but unaware state did not exist, which contradicted the evidence of asymptomatic patients who were not aware of the disease but was still infectious during the COVID-19 outbreak~\cite{rothe2020transmission}. These studies established a general framework that is applicable for analyzing the interplay between disease and information dynamics.
Nevertheless, few studies consider the coexistence of both local and global information dissemination, and global information such as mass media is often assumed to be independent of the network's disease dynamics. In a real-world setting, individuals may obtain information through their daily encounters (e.g., observing the symptoms and preventative measures from other travelers) and the mass media. Moreover, the information released by the mass media should depend on the actual disease patterns that are observable from the population. Besides, these early studies only focus on the equilibrium states for the disease, namely the disease-free state and the endemic state, while ignoring the terminal state of information and how the state of information may drive the equilibrium state of the infectious diseases. 


In this study, we develop a multiplex disease model that is suitable for modeling the disease dynamics with local observations and global information to bridge the gaps mentioned above. The multiplex model consists of the disease layer with a susceptible-infected-susceptible process (SIS), the local observation layer with an unaware-aware process (UA), and a global information node that disseminates information to individual nodes. The strength of the information depends on the states of the disease layer. We introduce the heterogeneous mean-field model (HMF)~\cite{pastor2001epidemic} to capture the collective disease and information dynamics of nodes with the same degree. By analyzing the theoretical properties of the UA-SIS model, we identify three critical equilibrium states that may emerge depending on the percolation strength of the disease and the information: the disease and awareness free equilibrium (DAFE), the disease-free equilibrium with awareness (DFE-A) and the endemic state. In particular, the DFE-A state corresponds to the case where the local information is strong enough to shape a herd-immunity-alike pattern among the nodes and eventually suppressed the spread of the diseases. Furthermore, we also demonstrate the disease and information dynamics of the UA-SIS model in the realistic contact network built from mobility patterns of metro travelers. 

The rest of the study is organized as follows. Section two introduces mathematical notations. Section three describes the assumption and supporting evidence. Section
four gives an overview of the UA-SIS model. Section five presents the mathematical
formulation of the UA-SIS model, and section six delivers a theoretical analysis of the
stable states and conditions for the UA-SIS model. Section seven shows the numerical experiments and insights obtained from the model. Section eight concludes the study with a summary and key findings.

%% file: notation.tex
\section{notation}
The mathematical notation used in this study is summarized in Table~\ref{tab:notation}.
\begingroup
\begin{table}[H]
	\caption{Table of notation}
	\label{tab:notation}
	\centering
	\begin{tabular}{p{2cm}p{12cm}p{1cm}}
		\hline
		\textbf{Variables} & \textbf{Descriptions}\\
		\hline
		$k$ & Nodes with degree $k$\\
		$U_k$ & Percentage of nodes in state $U$ with degree $k$. U=unaware.\\
		$\theta_{A,k}$ & Percentage of nodes in state $A$ with degree $k$. A=aware.\\
		$S_k$ & Percentage of nodes in state $S$ with degree $k$. S=susceptible.\\
		$\theta_{I,k}$ & Percentage of nodes in state $I$ with degree $k$. I=infectious.\\
		$\beta_U$ & Transmission probability when a node is in $U$ state. \\
		$\beta_A$ & Transmission probability when a node is in $A$ state. $\beta_A<<\beta_U$. \\
		$\gamma_1$ & Recovery rate of the aware state, $\gamma_1\leq1$. \\
		$\gamma_2$ & Recovery rate of the given disease, $\gamma_2\leq1$. \\
		$p$ & Probability of a node changes from $U$ to $A$ by observing that one of its neighbors is in $A$ state.\\
		$\mathcal{N}(i)$ & The set of neighbor nodes of node $i$. \\
		\hline
	\end{tabular}
\end{table}
\endgroup

%% file: assumption.tex
\section{assumption}
The following assumptions are made to support the development of the UA-SIS model:
\begin{enumerate}
\itemsep0em 
    \item It is assumed that nodes of the same degree have the same behavior. So that we can make use of the HMF model, also known as the degree-based mean-field model, to understand UA-SIS dynamics.  
    \item The discussion is currently limited to HMF without degree correlation. Nevertheless, the framework discussed in this study can be easily adapted to account for correlated degree sequences. 
    \item It is considered that both UA and SIS processes have the same network structure for the numerical experiments. This is to model the scenario where commuters obtain local information by observing the symptoms and preventative measures of their encounters during travel. Various network structures in different layers will be considered in future work. 
\end{enumerate}

%% file: ua_sis.tex
\section{UA-SIS model}
\begin{figure}[H]
    \includegraphics[width=0.7\linewidth]{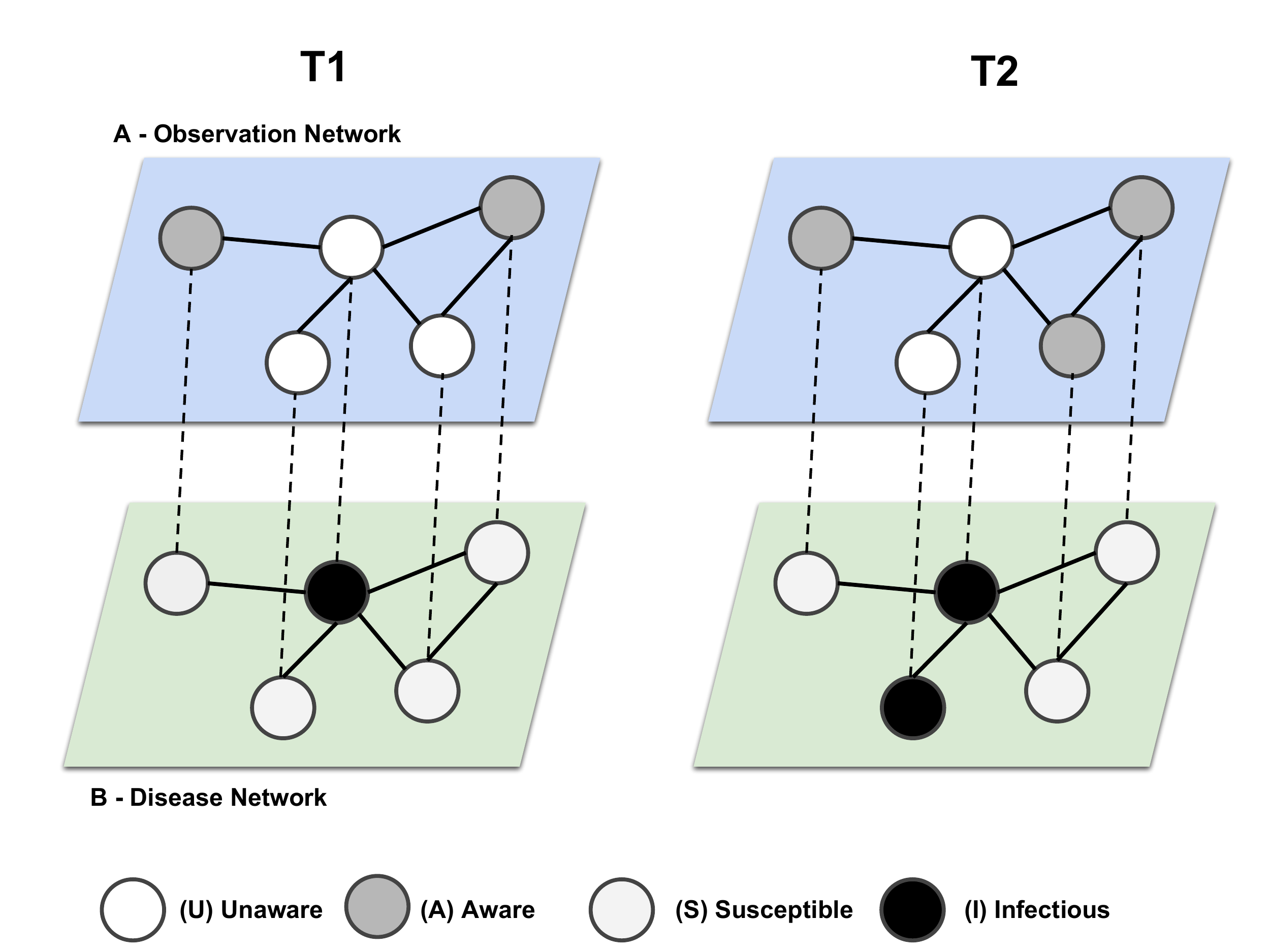}
    \caption{Illustration of the UA-SIS modeling process.}
    \medskip
    \small
    \justify
    (In the figure, the system has two layers of networks: the disease network and the observation network. The SIS process takes place on the disease network while the UA process is on the observation network. On the left (time=T1), the system starts with one infectious node, and two out of four of its neighbors are aware of his illness. On the right (time=T2), the system evolves over time, and one more node turns into A state. And the neighbor node who is still in U state is infected by the infectious node.)
    \label{fig:UA_SIS_fig}
\end{figure}

In the UA-SIS model, individuals are modeled as nodes, and their pairwise connections are captured by the edges. The UA-SIS model is capable of capturing the dynamics of three processes that take place simultaneously:
\begin{enumerate}
    \item The physical contagion process, where individuals get in contact with others and the disease may spread upon a contagion. 
    \item The observation process (local information), where individuals may observe the behavior of his/her neighbors, and accumulate knowledge of the diseases. Each valid observation will strengthen his understanding and increase the likelihood of an individual migrating from U state to A state. Meanwhile, individuals may move from A to U as time proceeds. This is known as the fading of memory. 
    \item The information gathering and dissemination process (global information). Unlike many other studies, where information dissemination is assumed to occur among neighbors only, this study considers the existence of a central system that gathers the disease information from the population and disseminates the compiled information back to the population within reach. 
\end{enumerate}
The interactions among these three processes can therefore be captured by a multiplex network with three levels, as shown in Figure~\ref{fig:UA_SIS_fig}.

In Figure~\ref{fig:UA_SIS_fig}, there are two correlated networks that represent disease dynamics and observation dynamics. In the observation network, nodes observe their neighbors and obtain local information about the disease information. If one of the neighbors is in A, then the node has a certain probability that it will evolve into A as well. As for the disease network, the classic SIS process is considered where the node is either in susceptible or infectious state. A susceptible node may turn into an infectious one upon physical contact with another infectious neighbor. However, since the two networks are inter-correlated and having the same topology, a susceptible node may not be infected by its infectious neighbors in the disease network if it is in the aware state in the observation network. This illustrates why using the SIS model itself may overestimate the outbreak scale of the disease and the necessity for including the information layer (observation as the case in our study). 

\begin{figure}[H]
    \includegraphics[width=\linewidth]{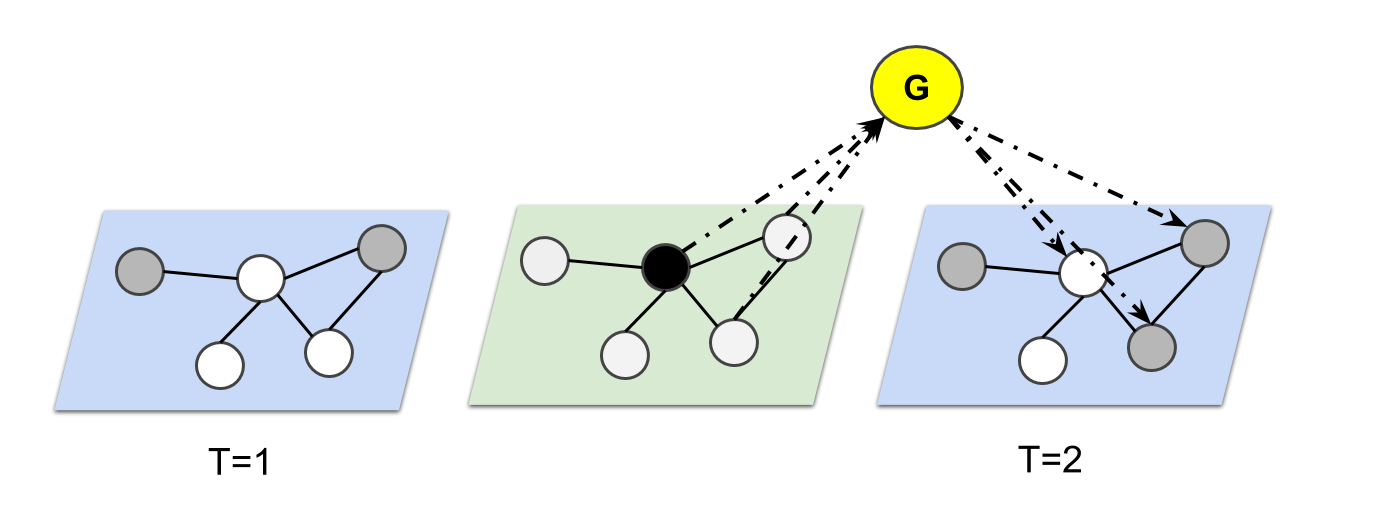}
    \caption{Illustration of the UA-SIS information dissemination process.}
    \medskip
    \small
    \justify
     (During the process, the central system (node G in yellow) obtains the disease information from its neighbors in the disease layer at time $T=1$, and the disseminates this information to the same set of nodes in the observation layer. This in combination with the observed information turns one of the nodes from U state to A state.)
    \label{fig:UA_SIS_info}
\end{figure}

Up to now, only local information dissemination upon physical contact is considered. However, one major source for the general public to obtain information is through online resources such as social media and news agencies. People may value this information differently, among which the most reliable source is the official data and news. However, it is in general difficult for any official media to understand the whole picture of the disease pattern. Consequently, the other important research question is to understand how different levels of information may affect the disease spreading dynamics over the disease network. This process is described in Figure~\ref{fig:UA_SIS_info}.

%% file: formulation.tex
\section{formulation}
The HMF method is used which assumes that nodes of same degree are homogeneous. Mathematically, the interplay among the three layers can be written as following. 

\subsection{Observation layer}
The observation layer describes the dynamics where each individual observes from his or her physical contacts and accumulating awareness of the disease states:
\begin{equation}
\begin{aligned}
    \frac{dU_k}{dt}=&-[1-(1-p)^{k\Theta_A(t)}+\lambda_g(t)+\beta_Uk\Theta_I(t)]U_k(t)\\
    &+\gamma_1 \theta_{A,k}(t)
\end{aligned}
    \label{eq:U2A}
\end{equation}
\begin{equation}
\begin{aligned}
    \frac{d\theta_{A,k}}{dt}=&[1-(1-p)^{k\Theta_A(t)}+\lambda_g(t)+\beta_Uk\Theta_I(t)]U_k(t)\\
    &-\gamma_1 \theta_{A,k}(t)
\end{aligned}
    \label{eq:A2U}
\end{equation}
where $\Theta_{A}(t)$ is the probability that an arbitrary neighbor of a node is in state $A$ at time $t$. $1-p$ denotes the probability that the node will remain in $U$, so that $(1-p)^{k\Theta_A(t)}$ gives the probability that node $i$ will remain in $U$ after observing all his neighbors in $A$. And $\lambda(g)$ captures the impact on the behavior due the information from central information node such as mass media, which will be explained in details in the following sections. As a consequence, the first term in equation~\ref{eq:U2A} refers to the proportion of nodes of degree $k$ that migrates from state $U$ to state $A$. Such a migration may take place upon observing the neighbors in A state, if the node itself is infected (captured by $\beta_Uk\Theta_I(t)$) or if the node is exposed to central information (e.g., the total number infections in the system) that motivates the change of behavior. And the second term describes the decreasing level of awareness over time so that people move from $A$ back to $U$ at the rate of $\gamma_1$. 

\subsection{Disease layer}
In disease networks, individuals are considered to be in one of the two states: susceptible (S) and infectious (I). In particular, for individual in S, the chance of being infected depends on if they are in A or U states, with different transmission coefficient $\beta_A$ and $\beta_U$ respectively. And the contagion dynamics between $S$ and $I$ states can be mathematically expressed as:


\begin{equation}
\begin{aligned}
\frac{dS_{k}}{dt}=&-\beta_Ak\Theta_I(t)S_{k}(t)\theta_{A,k}(t)-\beta_Uk\Theta_I(t)S_{k}(t)(1-\theta_{A,k}(t))\\
&+\gamma_2\theta_{I,k}(t)
\end{aligned}
\label{eq:sa2i}
\end{equation}

\begin{equation}
\begin{aligned}
\frac{d\theta_{I,k}}{dt}=&\beta_Ak\Theta_I(t)S_{k}(t)\theta_{A,k}(t)+\beta_Uk\Theta_I(t)S_{k}(t)(1-\theta_{A,k}(t))\\
&-\gamma_2\theta_{I,k}(t)
\end{aligned}
\label{eq:i2s}
\end{equation}

In the equations, $\Theta_{I}(t)$, similar to the notion of $\Theta_{A}(t)$, characterizes the probability that an arbitrary neighbor of a node is in $I$ state at time $t$. The first two terms in equations~\ref{eq:sa2i} captures the how susceptible individuals are infected under U and A states respectively, with $\theta_{A,k}(t)$ and $1-\theta_{A,k}(t)$ referring to the proportion of nodes in U and A states. And the third term in equation~\ref{eq:sa2i} suggests the recovery of individuals in $I$ state with the rate of $\gamma_2$. 

\subsection{Information layer}
We next introduce the equation for the information layer, where it is assumed that there is a central system that collects the information over the network. Typical examples of the information layer can be that people share their states and thoughts of the disease dynamics over the social networks, or the state mass media distribute the updated disease information from Centers for Disease Control and Prevention. 

    For the functionality of the information layer, we consider two types of information gathering schemes: the targeted information fetching and the random information fetching. It is considered that the central node can obtain the information over the disease network from the nodes that are adjacent to it, and then compile the information and send back to the same set of nodes in the observation network. And the expressions for the target and random information fetching are written as:
\begin{equation}[Target]
    \quad \lambda_g(t)=\kappa\sum_{i\in\mathcal{T}} \Theta_{I,\mathcal{T}}(t)
\end{equation}
\begin{equation}[Random]
    \quad \lambda_g(t)=\kappa\alpha\sum_k \theta_{I,k}(t)P(k)
\end{equation}
where $\Theta_{I,\mathcal{T}}(t)$ denotes the probability that an arbitrary selected node is infected within the target set $\mathcal{T}$. $\kappa$ is a discount factor that converts the total number of infectious nodes into the level of risk of the disease, and $\alpha$ is the ratio that accounts for the proportion of nodes that the central node has connection to randomly.

\subsection{Probability of a node in infectious state}
With the awareness, disease and information dynamics presented in the previous sections, we can now define the probability that a randomly selected neighbor of a node with degree $k$ is in $A$ or $I$ states, namely $\Theta_A(t)$ and $\Theta_I(t)$.

Since $\theta_{A,k}(t)$ gives the probability that a node of degree $k$ is in $A$ state at time $t$, we have:
\begin{equation}
    \Theta_A(t)=\frac{\sum_{k'}k'P(k')\theta_{A,k'}(t)}{<k>}
    \label{eq:adeg}
\end{equation}
where $<k>$ is the average degree of the network. Equation~\ref{eq:adeg} is the weighted expectation of nodes with degree $k$ where the degree distribution is characterized by $P(k)$. The underlying intuition behind this equation suggests that high degree node usually have a much higher chance of being in the infected states as compared to low degree nodes, so that such heterogeneity of node degree should be taken into consideration with the incorporation of the node degree distribution~\cite{pastor2015epidemic}. 

Similarly, we can formally written the probability that a randomly selected neighbor of a node with degree $k$ in $I$ state as:
\begin{equation}
    \Theta_I(t)=\frac{\sum_{k'}k'P(k')\theta_{I,k'}(t)}{<k>}
\end{equation}
$P(k)$ here is the same as that in equation~\ref{eq:adeg} as we assume the same network structure for the observation layer due to the consideration of physical contact. This consideration can be easily relaxed for the sake of other applications with different networks structure in the observation layer (e.g., observation through social network rather than physical contact network), where one can replace the $P(k)$ with a proper degree distribution that captures the network structure of the observation layer. 

%% file: analysis.tex
\section{System equilibrium and stability analysis}
SIS model is a well-studied epidemic model with two equilibrium states. One is the disease free equilibrium (DFE) where all individuals are susceptible. This can be analogous to that all individuals are in $U$ and $S$ states in our UA-SIS model, where such state is named as the disease and awareness free equilibrium (DAFE). The other equilibrium is known as the endemic equilibrium, where there will always be a proportion of nodes in infectious state, and the size of infectious population is equal to the size of the giant component in the graph. Finally, there is a special equilibrium point for the UA-SIS model, where the state is free from disease invasion but the awareness itself is permanent and strictly positive. This regime is named as the disease free equilibrium with awareness (DFE-A). 

\subsection{Disease and awareness free equilibrium}
We first discuss the first equilibrium state, the DAFE, so that when $t\rightarrow 0$ we have $\theta_{A,k}(t)=0$ and $\theta_{I,k}(t)=0$ for all $k$. Equivalently, this gives $U_k(t)=1$ and $S_{k}(t)=1$ in their corresponding layers respectively. This reduces equations~\ref{eq:A2U} and~\ref{eq:i2s} to
\begin{equation}
    \frac{d\theta_{A,k}}{dt}=1-(1-p)^{k\Theta_A(t)}+\lambda_g(t)-\gamma_1 \theta_{A,k}(t)
    \label{eq:alinear}
\end{equation}
\begin{equation}
    \frac{d\theta_{I,k}}{dt}=\frac{\sum_{k'}\beta_Ukk'P(k')\theta_{I,k'}(t)}{<k>}-\gamma_2 \theta_{I,k}(t)
    \label{eq:ilinear}
\end{equation}
In the neighborhood of DFE, we further know that $k\theta_A(t)\rightarrow 0$ and we should have 
\begin{equation}
    (1-p)^{k\Theta_A(t)}\approx 1+ln(1-p)k\Theta_A(t)
\end{equation}
so that we can rewrite equation~\ref{eq:alinear} as:
\begin{equation}
\begin{aligned}
    \frac{d\theta_{A,k}}{dt}&=-ln(1-p)k\Theta_A(t)+\lambda_g(t)-\gamma_1 \theta_{A,k}(t)\\
    &=-ln(1-p)k\Theta_A(t)+\kappa\alpha\sum_k \theta_{I,k}(t)P(k)-\gamma_1 \theta_{A,k}(t)
\end{aligned}
    \label{eq:alinearrelaxed}
\end{equation}
Let $C^{OD}$ be the correlation matrix between observation layer and disease layer, and $C^{OO}$ and $C^{DD}$ be the matrices for observation layer and disease layer respectively,where.
\begin{equation}
C^{OO}_{k_1,k_2}=-\frac{ln(1-p)k_1k_2P(k_2)}{<k>}
\label{coo}
\end{equation}
\begin{equation}
C^{OD}_{k_1,k_2}=\kappa\alpha P(k_2)
\end{equation}
\begin{equation}
C^{DD}_{k_1,k_2}=\frac{\beta_Uk_1k_2P(k_2)}{<k>}
\label{cod}
\end{equation}

We should therefore have 
\begin{equation}
    C=\begin{bmatrix}
    C^{OO} & C^{OD}  \\
    0 & C^{DD}  \\
\end{bmatrix}
\end{equation}

We can therefore write equations~\ref{eq:ilinear} and~\ref{eq:alinearrelaxed} as the following linear system:
\begin{equation}
    \frac{d\theta}{dt} = C\theta -\gamma . \theta
\end{equation}
where $\theta=[\theta_{A,1},...,\theta_{A,k_{max}},\theta_{I,1},...,\theta_{I,k_{max}}]^T$.

We should have the following proposition:
\begin{prop}
The DAFE of the UA-SIS system is asymptotically stable if the spectral radius of $\rho(C)<\gamma$. In other words, if the maximum eigenvalue of $C$ is smaller than $\gamma$, then the DFE is asymptotically stable. 
\end{prop}

Since $C$ is an upper-triangular matrix, we should have the maximum eigenvalue of C as
\begin{equation}
    \lambda=max(\lambda(C^{OO}),\lambda(C^{DD}))
\end{equation}

When $t\rightarrow \infty$, according to~\cite{moreno2002epidemic}, the disease transmission thresholds for observation layer and disease layer can be written as:

\begin{equation}
    \lambda(C^{OO})=\lambda_O=-ln(1-p)\frac{<k^2>}{<k>}
    \label{radius1}
\end{equation}
\begin{equation}
    \lambda(C^{DD})=\lambda_D=\beta_U\frac{<k^2>}{<k>}
    \label{radius2}
\end{equation}

We now describe the details to get \ref{radius1} and \ref{radius2}. In equations \ref{coo} and \ref{cod}, the constant items are respectively $-ln(1-p)/<k>$ and $\beta_{U}/<k>$. Except for the constant items, we focus on the matrix $M = [k_{i}k_{j}P(k_{j})]_{i,j}$. $M$ is a square matrix and can be expressed as the multiplication of a column vector and a row vector:
\begin{equation}
    M=\begin{bmatrix}
    k_{1} \\
    k_{2} \\
    \dots   \\
    k_{l}
\end{bmatrix}
\begin{bmatrix}
    k_{1}P(k_{1}) & k_{2}P(k_{2}) & \dots & k_{l}P(k_{l})
\end{bmatrix}
\end{equation}

We know for two matrices $A,B$, $rank(AB)\leq min\{rank(A),rank(B)\}$, so $rank(M)\leq 1$. In addition, all entries of $M$ are positive so $rank(M)\geq 1$. It follows that $rank(M)=1$. So 0 is an eigenvalue of $M$ with the geometric multiplicity as $l-1$. As geometric multiplicity of an eigenvalue is not smaller than its algebraic multiplicity, the algebraic multiplicity of 0 is $l-1$ or $l$. We also notice that the sum of all eigenvalues of $M$ is equal to the sum of all diagonal items in $M$, which is $\sum_{i=1}^{l}k_{i}k_{i}p(k_{i})=<k^{2}>>0$. Therefore, the algebraic multiplicity of 0 is $t-1$ and the other non-zero eigenvalue is $<k^{2}>$. So the spectral radius of $M$ is $<k^{2}>$ and we have equations~\ref{radius1} and~\ref{radius2}.

\begin{figure}[H]
    \centering
    \includegraphics[width=0.6\linewidth]{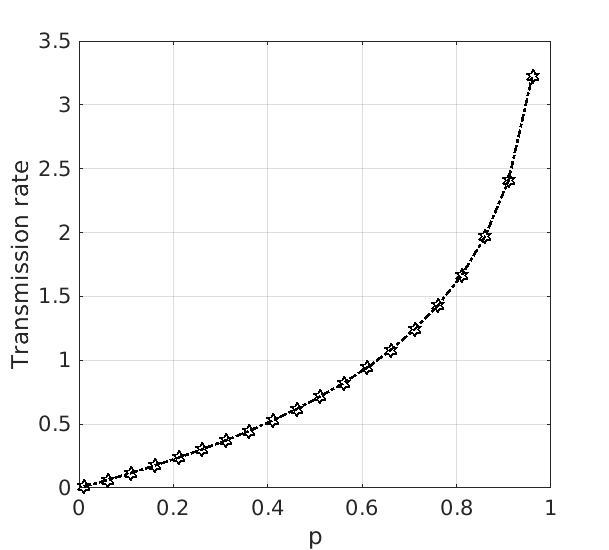}
    \caption{Spreading rate of awareness with the change of individual perception of disease}
    \label{fig:awareness}
\end{figure}

Following equation~\ref{radius1}, we immediately observe that the spreading spread of awareness decays exponential with the strength of individual perceptions of the disease, as shown in Figure~\ref{fig:awareness}

The second observation from the structure of $C$ is that the availability of global information does not affect the disease threshold in a network. That is, $\lambda_g$ does not determine the value of $\lambda_O$ and $\lambda_D$, and hence $\lambda_C$. This observation may sound counter-intuitive at first glance. However, when disease is approaching DFE, the value of $\lambda_g$ is nearly zero as there are barely any infected people in the network. As a consequence, it is always of lower order as compared to the personal awareness of the disease, which plays a major role in the spreading process. The individual perception level directly gives the duration that the awareness may persist, and therefore how likely that people may stay in a safer state. 

\subsection{Disease free equilibrium with awareness}
By observing the system equations, there are actually two different DFEs rather than one for basic SIS model: the DAFE, which is disease and awareness free equilibrium, and the DFE-A, which is the disease free equilibrium with positive awareness population. The DAFE state is discussed in the previous section. As for DFE-A, the equilibrium point of interest is $\theta_{I,k}=0$, $\theta_{A,k}\geq 0$. 

To calculate the equilibrium point of interest, we first note that at DFE-A, $\Theta_I=0$ but $\Theta_A>0$. By taking $\frac{d\theta_{A,k}}{dt}=0$ at DFE-A with above information, we should have
\begin{equation}
    [1-(1-p)^{k\Theta_A}](1-\theta_{A,k})-\gamma_1 \theta_{A,k}=0
\end{equation}
and this yields
\begin{equation}
    \theta_{A,k}=\frac{1-(1-p)^{k\Theta_A}}{1-(1-p)^{k\Theta_A}+\gamma_1}
\end{equation}
Introducing this equation to equation~\ref{eq:adeg} for $\Theta_A(t)$, we get a self-consistent equation for $\Theta_A$ as
\begin{equation}
    \Theta_A=\frac{1}{<k>}\sum_{k'} kP(k')\frac{1-(1-p)^{k'\Theta_A}}{1-(1-p)^{k'\Theta_A}+\gamma_1}
\end{equation}

where $0$ is an trivial solution that corresponds to the DAFE equilibrium. The function 
\begin{equation}
    f(\Theta_A)=\frac{1}{<k>}\sum_{k'} kP(k')\frac{1-(1-p)^{k'\Theta_A}}{1-(1-p)^{k'\Theta_A}+\gamma_1}-\Theta_A
\end{equation}
is a concave function, where we have
\begin{equation}
    \frac{1-(1-p)^{k'\Theta_A}}{1-(1-p)^{k'\Theta_A}+\gamma_1}<1
\end{equation}
This suggest that $f(1)<0$ and $\frac{df(x)}{dx}|x=0>0$. As a consequence, $\Theta_A$ admits a positive solution within the interval $(0,1)$. Without loss of generality, consider the DFE-A solution being $(\mu,1-\mu,1,0)$ where 
\begin{equation}
    \mu_k= \frac{\gamma_1}{1-(1-p)^{k\Theta_A^*}+\gamma_1}
\end{equation}
And the solution with positive awareness is stable as long as $\rho (C)>\gamma_1$, which can be derived based on our analysis for the stability of the DAFE. Now we know that $\frac{\gamma_1}{1+\gamma_1}\leq\mu_k\leq 1$. More importantly, for nodes with higher degree, $\mu_k\rightarrow \frac{\gamma_1}{1+\gamma_1}$. Meanwhile, the higher the $p$  is, the lower the $\mu_k$ will be. And the value of $\mu_k$ is sensitive to $\gamma_1$ value, which is the fading rate of memory. If it takes longer for people to forget the impacts of the diseases, then the $\gamma_1$ value should be smaller and there will be fewer people in $U$ state.

We now linearize the UA-SIS system at DFE-A as:
\begin{equation}
    \frac{d\theta_{A,k}}{dt}=[1-(1-p)^{k\Theta_A(t)}+\lambda_g(t)+\beta_Uk\Theta_{I}(t)]\mu_k-\gamma_1 \theta_{A,k}(t)
\end{equation}
\begin{equation}
\frac{d\theta_{I,k}}{dt}=\beta_Ak\Theta_I(t)(1-\mu_k)+\beta_Uk\Theta_I(t)\mu_k-\gamma_2\theta_{I,k}(t)
\end{equation}
Rearranging the right hand side gives:
\begin{equation}
    \frac{d\theta_{I,k}}{dt}=\beta_Ak\Theta_I(t)+(\beta_U-\beta_A)\mu_kk\Theta_I(t)-\gamma_2\theta_{I,k}(t)
\end{equation}
To ensure that the DFE-A is a.s.s, we just need to ensure that $I=0$ is stable solution, which is equivalent to that 
\begin{equation}
    \lambda_{DFE-A}=\beta_A\frac{<k^2>}{<k>}+(\beta_U-\beta_A)\frac{<\mu_k k^2>}{<k>}<\gamma_2
    \label{eq:dfea}
\end{equation}

If we consider that $\beta_A=0$ or $\beta_A<<\beta_U$, which is equivalent to that those people who are aware of the disease will be totally vaccinated or quarantined, we have that $\lambda_D^{DFE-A}$ is proportional to $(\beta_U-\beta_A)\frac{<\mu_k k^2>}{<k>}$. This indicates that the local information contributes significantly to lowering the disease threshold as compared to the state when there is no information available. But the marginal gain will be considerably weaker as we keep increasing the value of $p$.

Based on these analysis, there will be two conditions for the disease to reach DFE-A state. The first is the trivial case, where the disease system itself will stay in the DFE state and the aware population is positive:
\begin{equation}
    \lambda_O>\gamma_1, \lambda_D<\gamma_2
\end{equation}

The second case is more interesting where the disease itself may initially land a local outbreak without significant level of awareness among the nodes. Then the awareness is built among the nodes and the strength of the disease is weakened by the spread of awareness so that the previously disease outbreaks will eventually be suppressed and eventually eliminated. Mathematically, this requires the following condition to be satisfied:
\begin{equation}
    \lambda_{DFE-A}<\gamma_2,\lambda_O>\gamma_1,\lambda_D>\gamma_2
\end{equation}
where both local outbreaks for disease and awareness are possible due to the threshold values for the disease and observation layers separately. But the joint eigenvalue is less than the critical threshold so that the disease will eventually be eliminated. 

\subsection{Size of endemic state}
We are not only interested in DFE of the diseases, but also would like to explore how local observation and global information may affect the speed of the infectious diseases, and consequently the size of the outbreak (endemic state). This motivates us to conduct further analysis. 

When $\lambda_D^{DFE-A}>\gamma_2$ and $\lambda_O>\gamma_1$, the disease will eventually reach the endemic state. Following~\cite{pastor2015epidemic}, we first calculate the size of the endemic disease. At endemic, we should have the equilibrium point being $(\mu,1-\mu,s,1-s)$ with $1\geq\mu>0$ and $1\geq s>0$. As a consequence, we should have
\begin{equation}
\begin{aligned}
    \frac{d\theta_{I,k}}{dt}=&(\beta_Ak\Theta_I(t)+(\beta_U-\beta_A)\mu_kk\Theta_I(t))(1-\theta_{I,k}(t))\\
    &-\gamma_2\theta_{I,k}(t)
\end{aligned}
\end{equation}

At endemic state, we should have $\frac{d\theta_{I,k}}{dt}=0$, so that 
\begin{equation}
    \gamma_2\theta_{I,k}=[\beta_A+(\beta_U-\beta_A)\mu_k](1-\theta_{I,k})k\Theta_I
\end{equation}
\begin{equation}
    \theta_{I,k}=\frac{\alpha_k k\Theta_I}{\gamma_2+\alpha_k k\Theta_I} 
    \label{eq:newtheta}
\end{equation}
where $\alpha_k=\beta_A+(\beta_U-\beta_A)\mu_k$. 
And
\begin{equation}
    \Theta_I=\frac{\sum_{k'} k'P(k')\theta_{I,k'}}{<k>}
\end{equation}
\begin{equation}
    \Theta_I=\frac{1}{<k>}\sum_{k'} k'P(k')\frac{\alpha_k' k'\Theta_I}{\gamma_2+\alpha_k' k'\Theta_I} 
    \label{eq:newtheTa}
\end{equation}
Following the same analysis as for $\Theta_A$ for DFE-A, we know that there will be a positive $\Theta_I$ in the interval $(0,1)$ which satisfies above equation. 

Meanwhile, for $\theta_{A,k}$, by setting 
\begin{equation}
        [1-(1-p)^{k\Theta_A(t)}+\lambda_g(t)+\beta_Uk\Theta_{I}(t)](1-\theta_{A,k}(t))=\gamma_1 \theta_{A,k}(t)
\end{equation}
\begin{equation}
        \theta_{A,k}=1-\frac{\gamma_1}{1-(1-p)^{k\Theta_A}+\lambda_g+\beta_Uk\Theta_{I}+\gamma_1}
\end{equation}

Let $G(k)=1-(1-p)^{k\Theta_A}+\lambda_g$, we have
\begin{equation}
    \theta_{A,k}=1-\frac{\gamma_1}{G(k)+\gamma_1+\beta_Uk\Theta_I}
    \label{eq:uk}
\end{equation}
\begin{equation}
    \mu_k=\frac{\gamma_1}{G(k)+\gamma_1+\beta_Uk\Theta_I}
    \label{eq:uk}
\end{equation}

Finally, we have
\begin{equation}
    \Theta_A=\frac{1}{<k>}\sum_{k'} k'P(k')(1-\mu_{k'})
\end{equation}
\begin{equation}
    \Theta_I=\frac{1}{<k>}\sum_{k'} k'P(k')(1-\frac{\gamma_2}{\gamma_2+(\beta_A+(\beta_U-\beta_A)\mu_{k'}) k'\Theta_I})
\end{equation}

Based on the equations, we observe that, for high degree nodes, the value of $u_k$ is dominated by $\beta_U\Theta_I$ rather than the strength of the personal awareness. But for nodes with low degree, the strength of information dictates the value of $\mu_k$. This suggests that at endemic state, high degree nodes are less prone to infections, while low degree nodes are more vulnerable to the risk of infectious diseases.

%% file: results.tex
\section{Numerical experiments}
\subsection{Experiment setting}
The ODE45 solver in MATLAB is used to simulate the UA-SIS model described in this study. And we present the dynamics of the UA-SIS model over two types of networks. The first network is the scale-free network, which is commonly observed network structure in the real-world such as the World Wide Web network and the social network~\cite{barabasi2003scale}. The degree distribution for the scale-free network follows:
\begin{equation}
    P(k)\propto k^{-\gamma}
\end{equation}

The other type of network we consider here is the realistic contact network observed in the transportation system. In particular, Qian et al.~\cite{qian2020scaling} developed the degree distribution for the contact networks in the metro system based on the smart card transaction data from three cities. The metro contact network (MCN) serves as an ideal candidate network to understand the change of behavior upon observing the states of the neighbors during travel. The degree distribution of the MCN follows:
\begin{equation}
P(k)=\sum_{i=1}^N \frac{(Mw_i)^ke^{-Mw_i}}{k!}P(t_i)
\label{eq:generation_model}
\end{equation}
where $M=2\sum_i^N \alpha t_i^{\gamma_t} (N-1)$ measures the number of contacts among $N$ travelers and $P(t_i)$ is the distribution of travel time in metro systems. The shape of MCN is controlled by two parameters: $\gamma_t$ which measures the similarity of mobility pattern among travelers and $\alpha$ captures the structural property of the metro networks. We set $\gamma_t=0.652$ and $\alpha=0.004$ following the parameters for the Shanghai metro network in~\cite{qian2020scaling}. For both type of networks, the results are obtained by simulating the disease dynamics over the network with $10^3$ nodes. And the scale-free network is generated so that the average node degree $<k>$ is the same as that of MCN for fair comparison. 

Finally, for all the experiments, we set $\gamma_1=\gamma_2=0.1$ and $\kappa=\alpha=0.05$ unless otherwise specified. 

\subsection{Results}
\begin{figure}[H]
    \centering
    \subfloat[Results for scale-free network\label{fig:sf_IA}]{\includegraphics[width=0.7\linewidth]{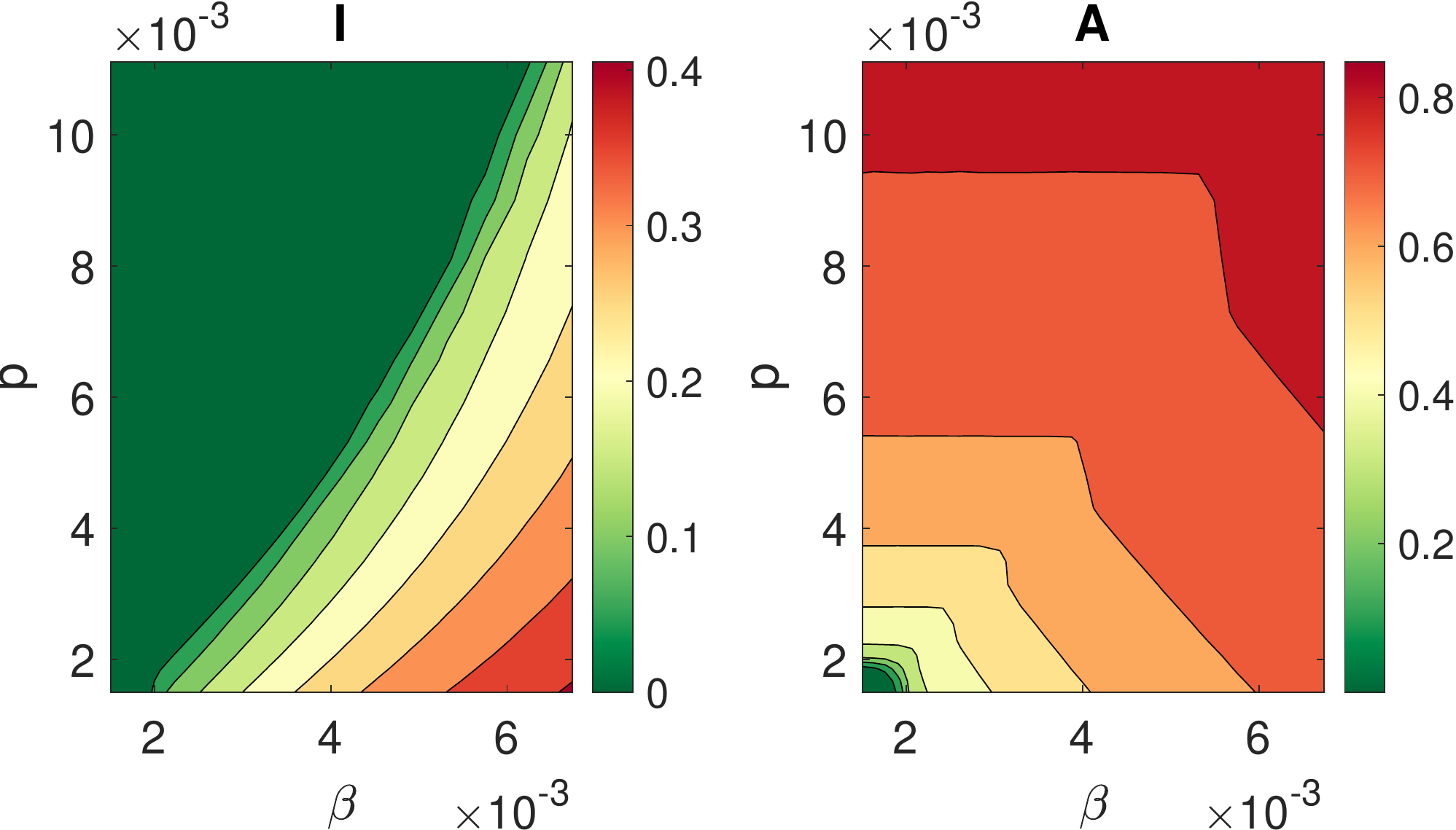}}\\
    \subfloat[Results for metro MCN\label{fig:mcn_IA}]{\includegraphics[width=0.7\linewidth]{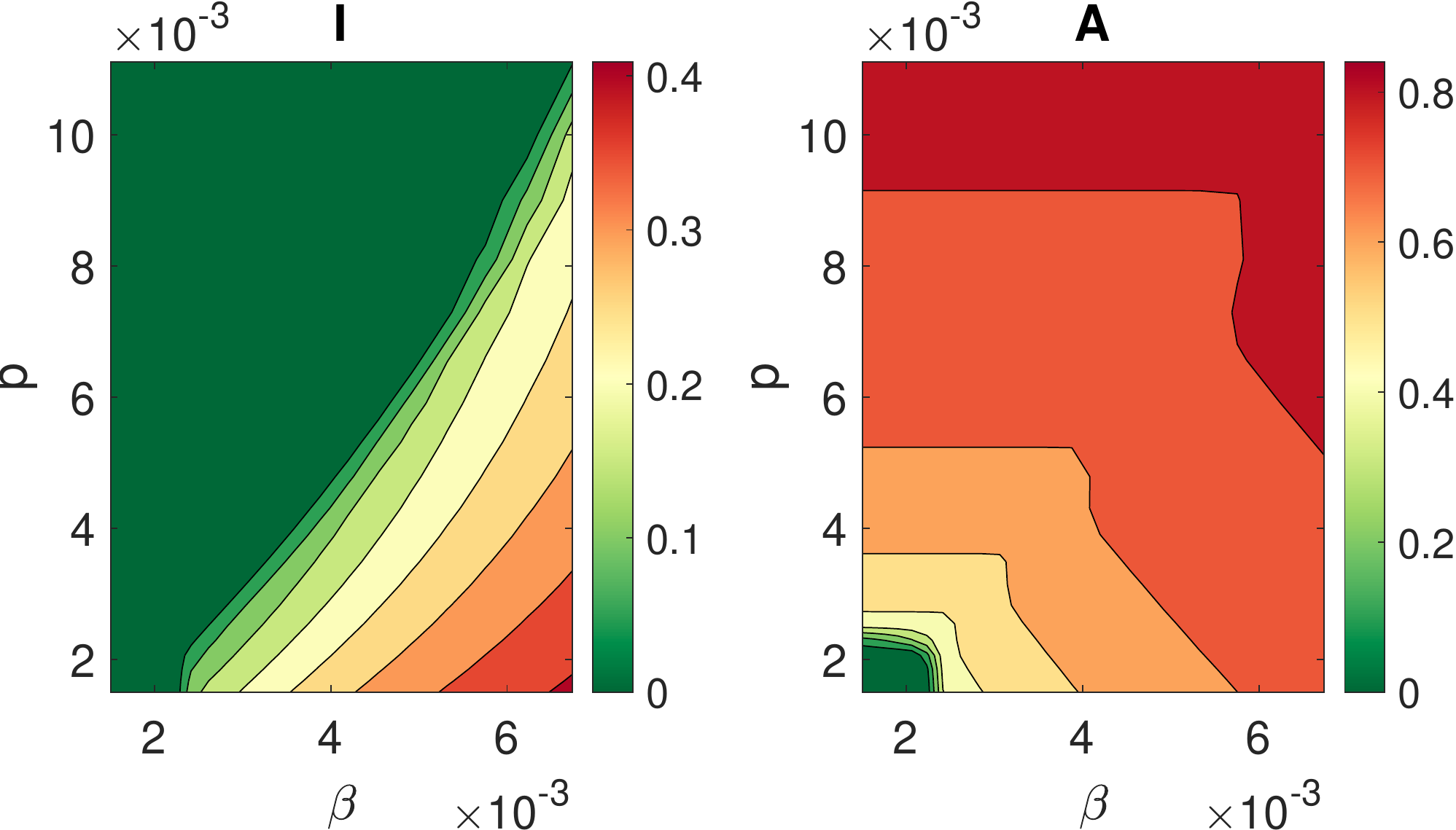}}
    \caption{Infectious population (left) and aware population (right) with varying $\beta$ and $p$ with $t\rightarrow\infty$.}
    \label{fig:compare_beta_p}
\end{figure}

Figure~\ref{fig:compare_beta_p} shows how varying $\beta$ and $p$ values may affect the final infectious and awareness population. The figure depicts the state transition among the three equilibrium states with different $\beta$ and $p$ values. In particular, when $\beta<0.002$, the DFE (I=0) can be achieved without the help of information dissemination and local awareness for both networks. Meanwhile, when $\beta<0.0025$ and $p<0.002$ in MCN, the strength of the information percolation is not strong enough to promote a local awareness outbreak over the observation network. And similar observation can be made for the scale-free network. Under such circumstances, the system is in DAFE state with disease and awareness population (I and A) being 0. When $\beta$ increases beyond $0.0025$ for MCN, the local disease outbreak starts to emerge but may be eliminated with a higher $p$ value. Such a state corresponds to the $DFE-A$ equilibrium, where the awareness is strictly positive, but the disease is eradicated. And the contour plot suggests the relationship between $\beta$ and $p$ to achieve DFE-A follows a convex function. As such, the increase in the pace of local awareness dissemination is expected to be greater than the increase in disease intensity so that a DFE can be achieved. Finally, when the disease strength exceeds $\lambda_{DFE-A}$, the system will transit into an endemic state, and both disease and awareness are permanent in the population. This corresponds to the areas with both I and A population are greater than zero in the contour plots. By examining the differences of the disease dynamics between the scale-free network and MCN as in Figure~\ref{fig:compare_beta_p}, we can tell that the area of the regions for DAFE and DFE-A are both greater for MCN than those for the scale-free network. This indicates that the MCN is more resilient to the threat of infectious diseases as compared to the scale-free network. This can be explained from their degree distribution where MCN decays faster than the scale-free network and presents hubs. These translate into lower $<k^2>$ for MCN so that the disease and awareness thresholds are shown in equations~\ref{radius1} and~\ref{radius2} are smaller since all other parameters are the same. Consequently, the real-world contact networks in the transportation system are less vulnerable than the scale-free network given its structural properties, where the latter is the network of interests in most network epidemiology studies. However, cautions still need to be exercised as the level of resilience of MCN is not significantly higher than that of the scale-free network, since $<k^2>$ will still diverge with increasing network sizes as discussed in~\cite{qian2020scaling}. On seeing the major differences between the MCN and the scale-free network, we focus exclusively on MCN for the following analyses.

\begin{figure}[!htbp]
     \subfloat[Terminal I population with different $\beta$ and $p$]{\includegraphics[width=.5\linewidth]{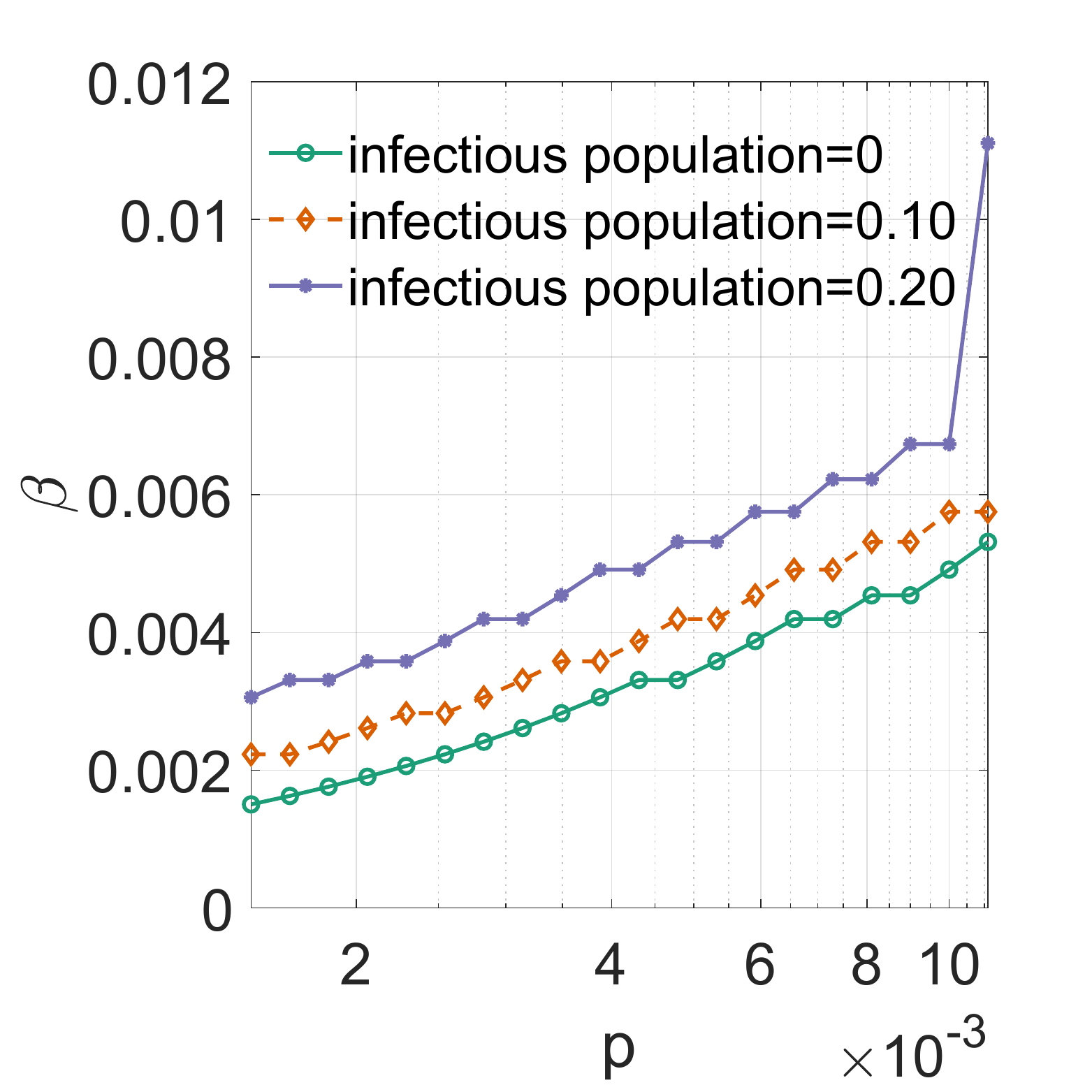}\label{<figure1>}}
     \subfloat[Terminal A population with different $\beta$ and $p$ ]{\includegraphics[width=.5\linewidth]{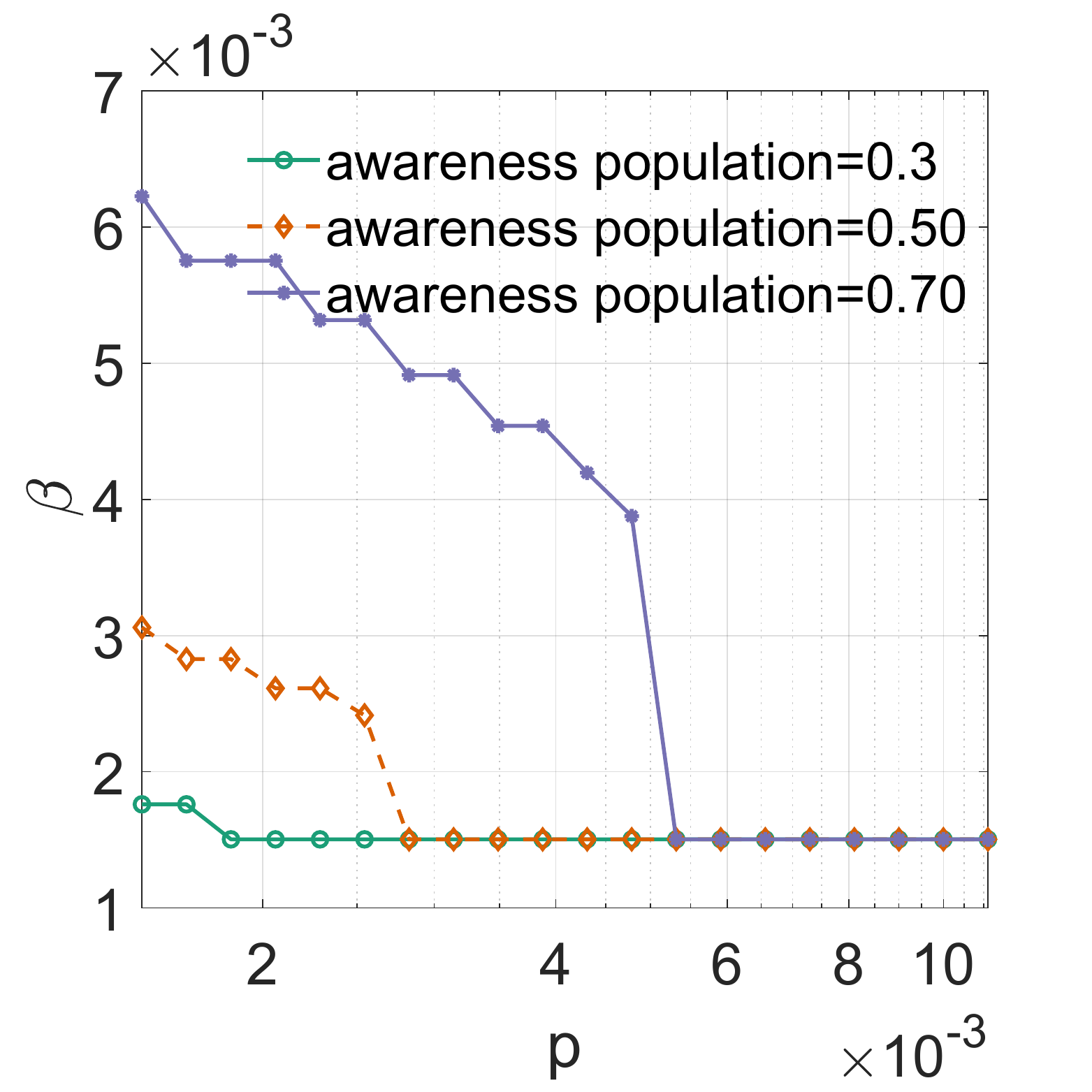}\label{<figure2>}}\\        
     \subfloat[Change of terminal I population with respect to $p$]{\includegraphics[width=.5\linewidth]{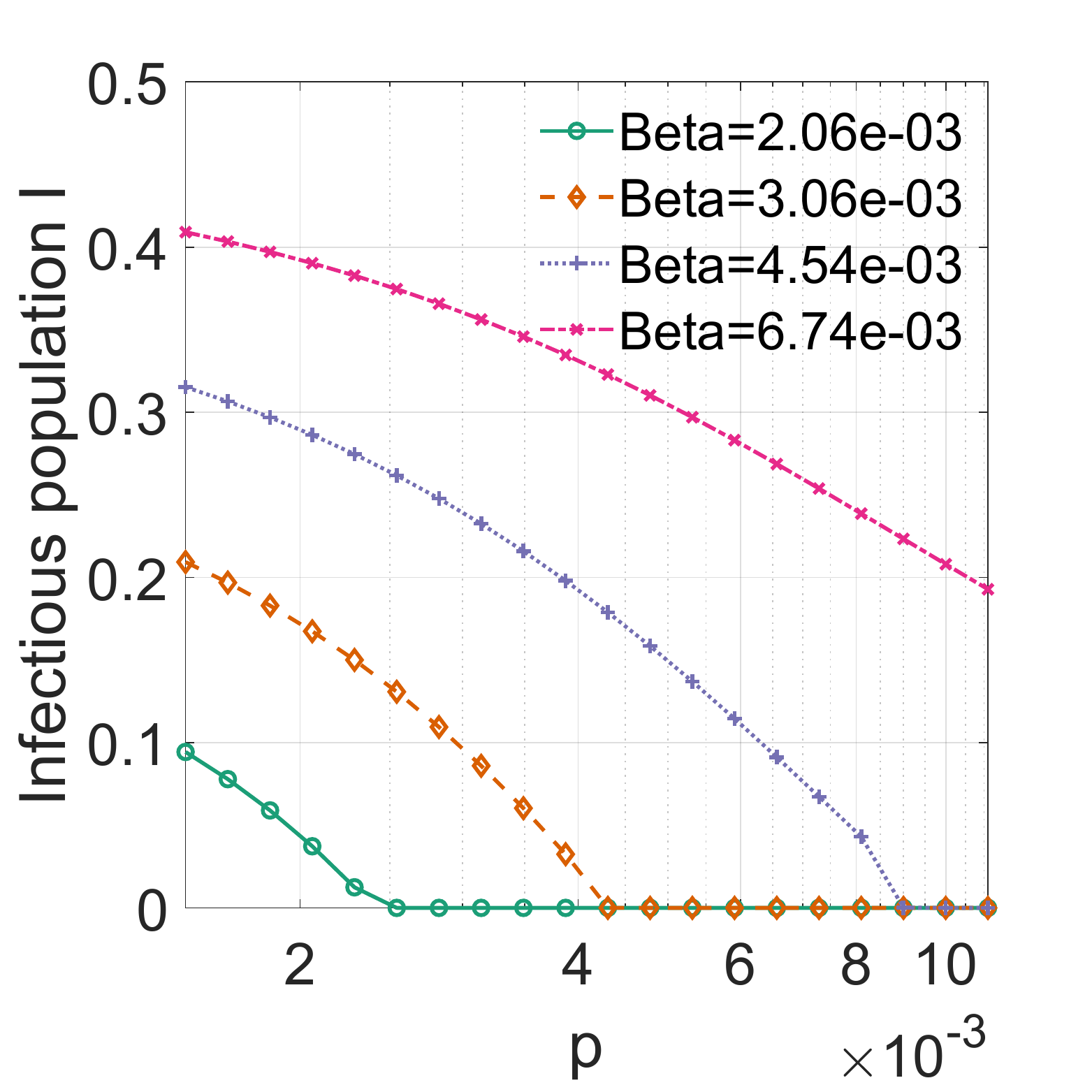}\label{<figure2>}}          
     \subfloat[Change of terminal A population with respect to $p$]{\includegraphics[width=.5\linewidth]{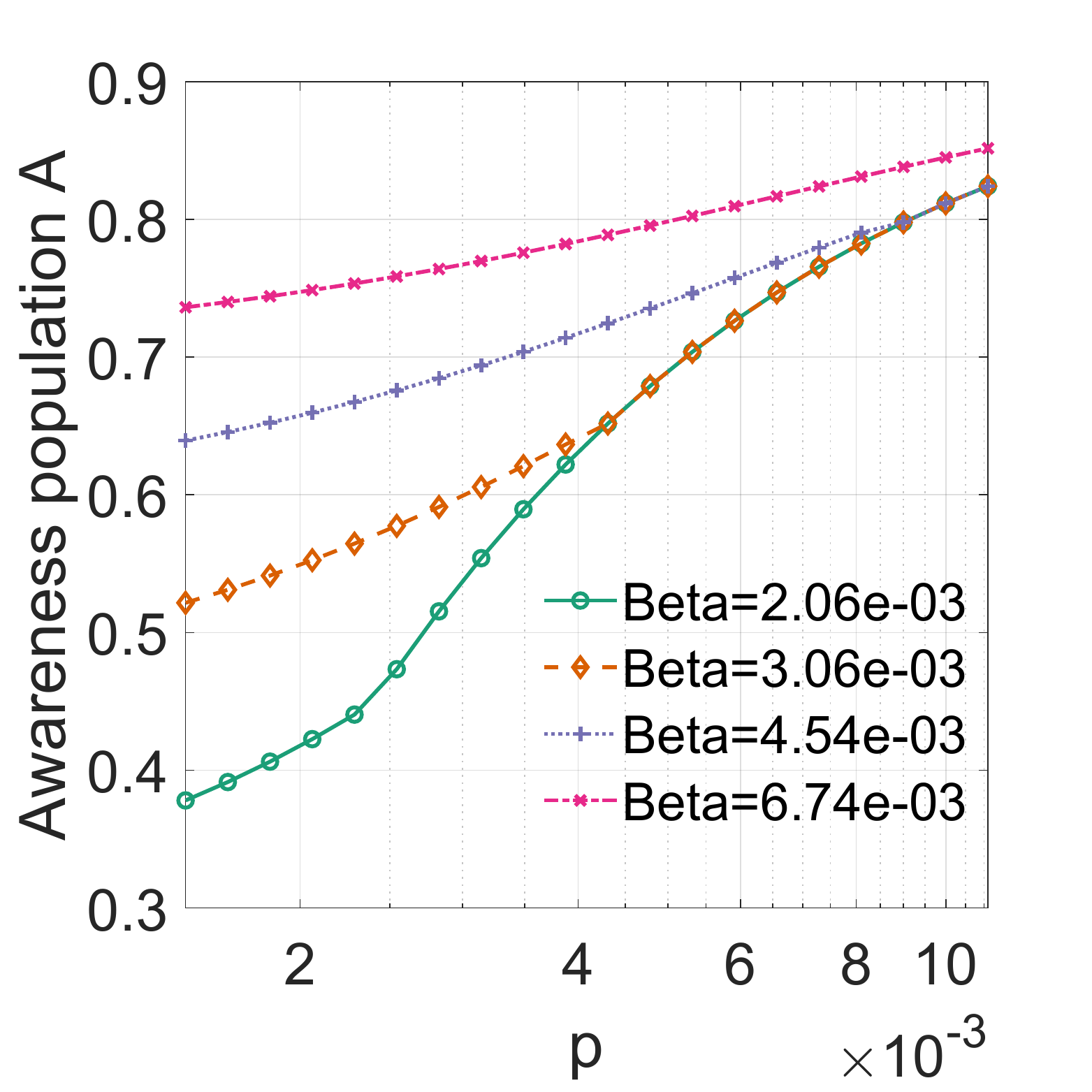}\label{<figure2>}}
    \caption{The impact of local information on disease spreading dynamics}
    \label{fig:localinfo}
\end{figure}

Figure~\ref{fig:localinfo} reveals the effect of local information transmission on spreading dynamics. The line corresponding to I=0 in Figure~\ref{fig:localinfo}(a) presents that local information efficiently suppresses disease outbreak, for the threshold increases (with increasing $\beta$) as p increases. This conclusion can also be drawn via equation~\ref{eq:dfea}, where $\mu_k$ is less than 1. For reaching a higher infectious population ($I=0.1$ and $0.2$), increasing individual awareness is observed to be less effective. Figure~\ref{fig:localinfo}(b) shows that increasing transmission probability $\beta$ lowers the threshold for awareness outbreak, as the higher p value is required to retain the same level of terminal A population. An interesting observation is the existence of “threshold drop”, which happens at around $p=0.0055$  for awareness population=0.7, $p=0.0037$ for awareness population=0.5 and $0.0015$ for awareness = 0.3. When $p$ increases from a lower value to above the observed threshold, the local information threshold drops rapidly. After the “threshold drop”, even a small percent of infection will lead to the local awareness spreading very quickly. This finding demonstrates the essential role of local information transmission related to an infectious disease outbreak. On the other hand, Figures~\ref{fig:localinfo}(c)-(d) show the slices of the corresponding parts in Figure~\ref{fig:compare_beta_p} and present that local information transmission reduces the limiting size of the infectious population and increases the limiting size of the aware population. This asymmetrical phenomenon is consistent with the results shown in Figures~\ref{fig:localinfo}(a)-(b). Finally, as $\beta$ increases, $p$ is found to have diminishing impacts on the final infectious and aware population which can be seen in Figures~\ref{fig:localinfo}(c)-(d). In these cases, $\beta$ dominates the system dynamics for both infectious and aware population, and this observation is consistent with the convex relationship between $\beta$ and $p$ for I population as can be seen in Figure~\ref{fig:compare_beta_p}

\begin{figure}[H]
\centering
    \includegraphics[width=1\linewidth]{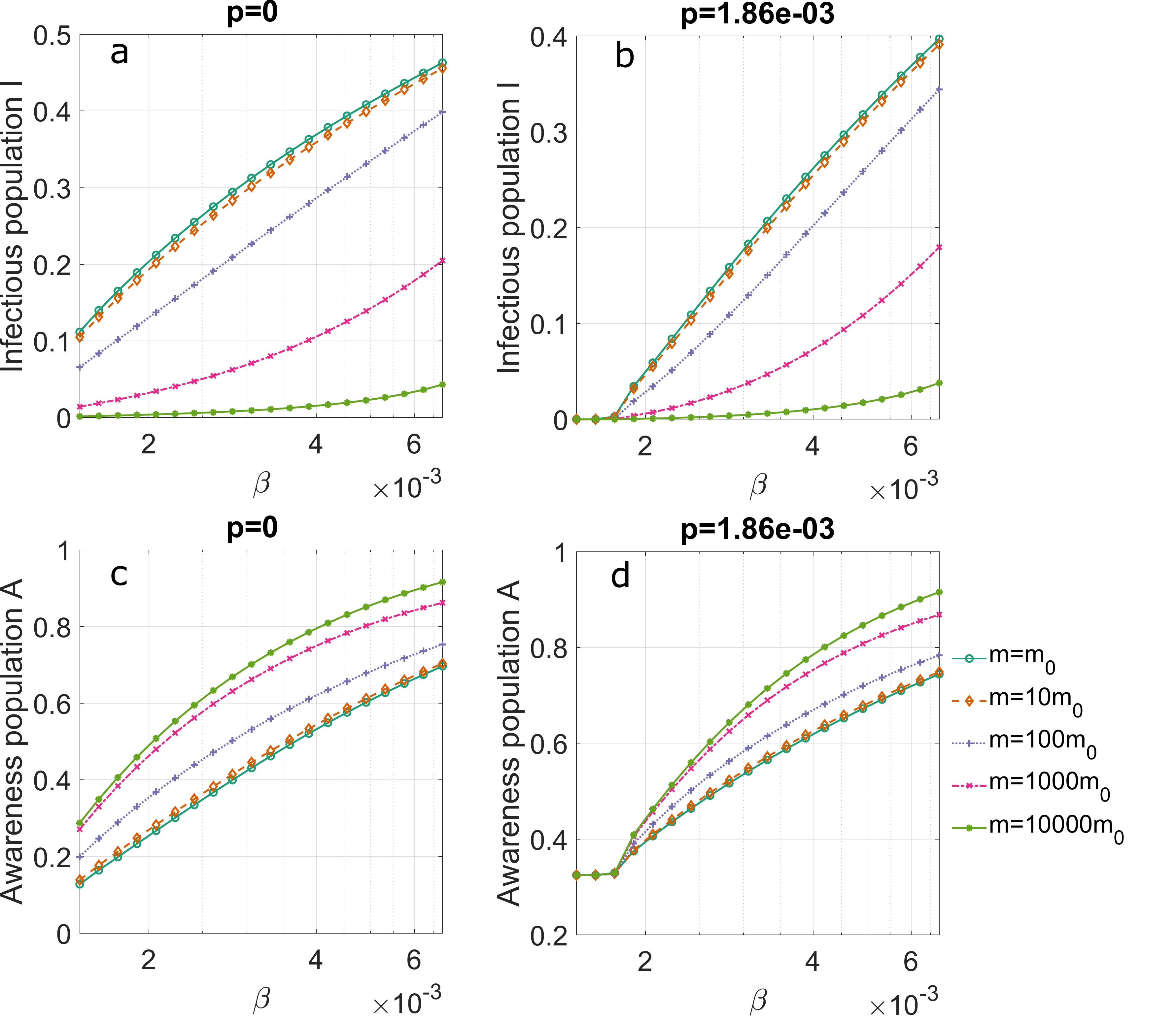}
    \caption{The impact of global information (random) on disease spreading dynamics}
    \label{fig:globalinfo}
\end{figure}

Figure~\ref{fig:globalinfo} reveals the role of global information by fixing the value of $p$. It can be directly verified from Figures~\ref{fig:globalinfo}(b) that global information has no impact on the threshold for disease-free equilibrium, as the infectious population will only be greater than 0 once a certain $\beta$ value is reached. Similarly, the global information will also not change the awareness threshold if there is no infectious population since the awareness population starts to increase as the infectious population is greater than 0 as demonstrated in Figure~\ref{fig:globalinfo}(d). These two results illustrate the dependency of global information on the disease states, as media agencies release global information by monitoring the progress of the disease dynamics in a reactive manner. After an infectious disease breaks out, only very strong global information is able to restrain disease efficaciously (e.g., when global information strength increased by $10^3$ and $10^4$ times). Similar observations can also be identified from Figures~\ref{fig:globalinfo}(a) and (c). And the key difference is that the combinations of $p$ with initial $\beta$ values results in endemic states in Figures~\ref{fig:globalinfo}(a) and (c) and DFE-A states in Figures~\ref{fig:globalinfo}(b) and (d). In conclusion, we observe both local and global information are powerful tools to suppress the spreading of infectious diseases. Local information is found to be more effective than the global information as it may percolate across the networks even without the presence of the disease outbreak. This can contribute to building a barrier against infectious diseases in a proactive manner. On the other hand, global information reacts upon the state of the infectious population, and can only help mitigate the scale of an outbreak after the disease invades the population. This can be understood as the case where mass media may not catch up on the progress of the diseases if it has not become a major threat to the general public. 




\begin{figure}[H]
    \centering
     \subfloat[DAFE]{\includegraphics[width=0.5\linewidth]{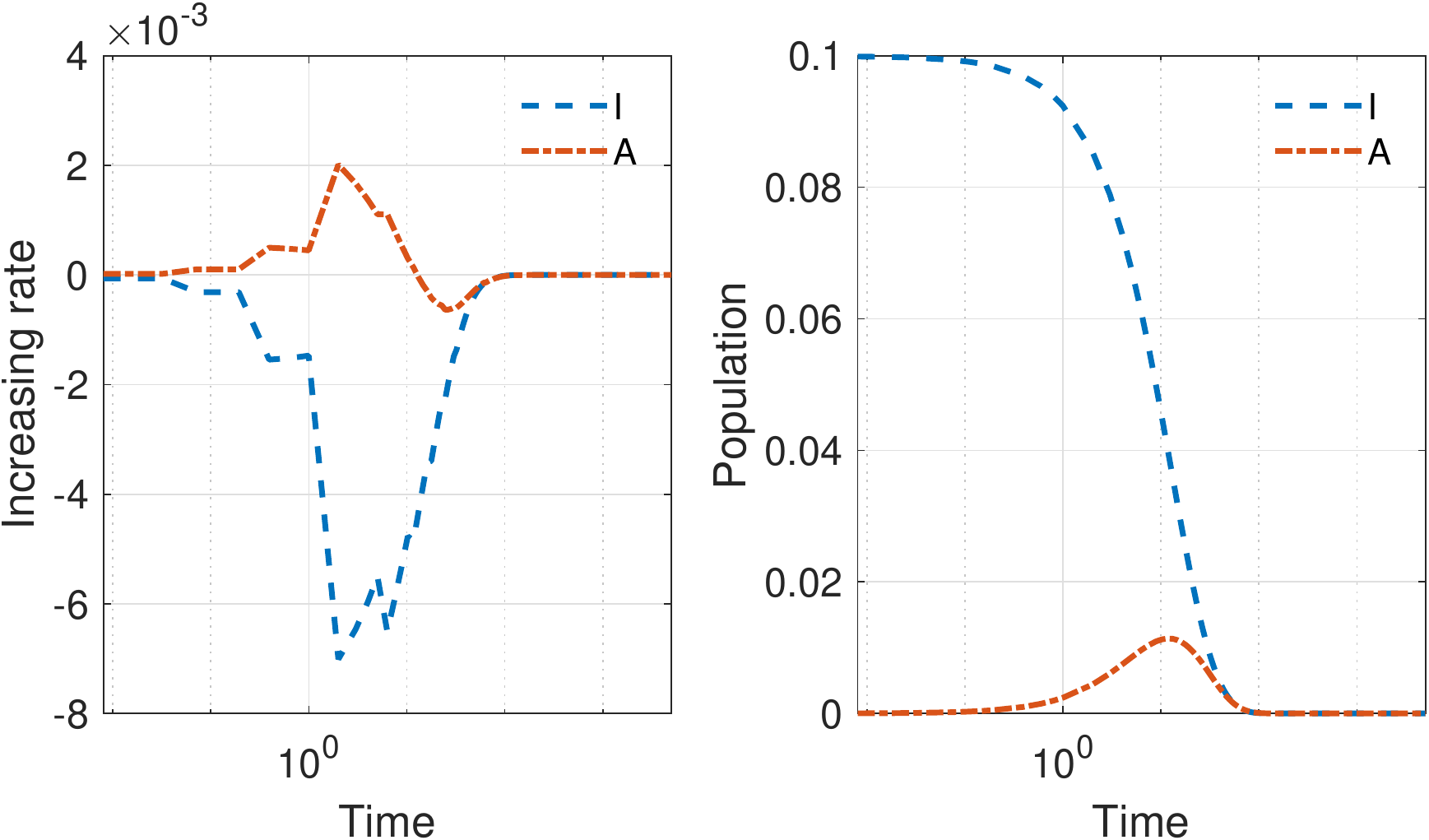}\label{<figure1>}}
     \subfloat[DFE-A]{\includegraphics[width=0.5\linewidth]{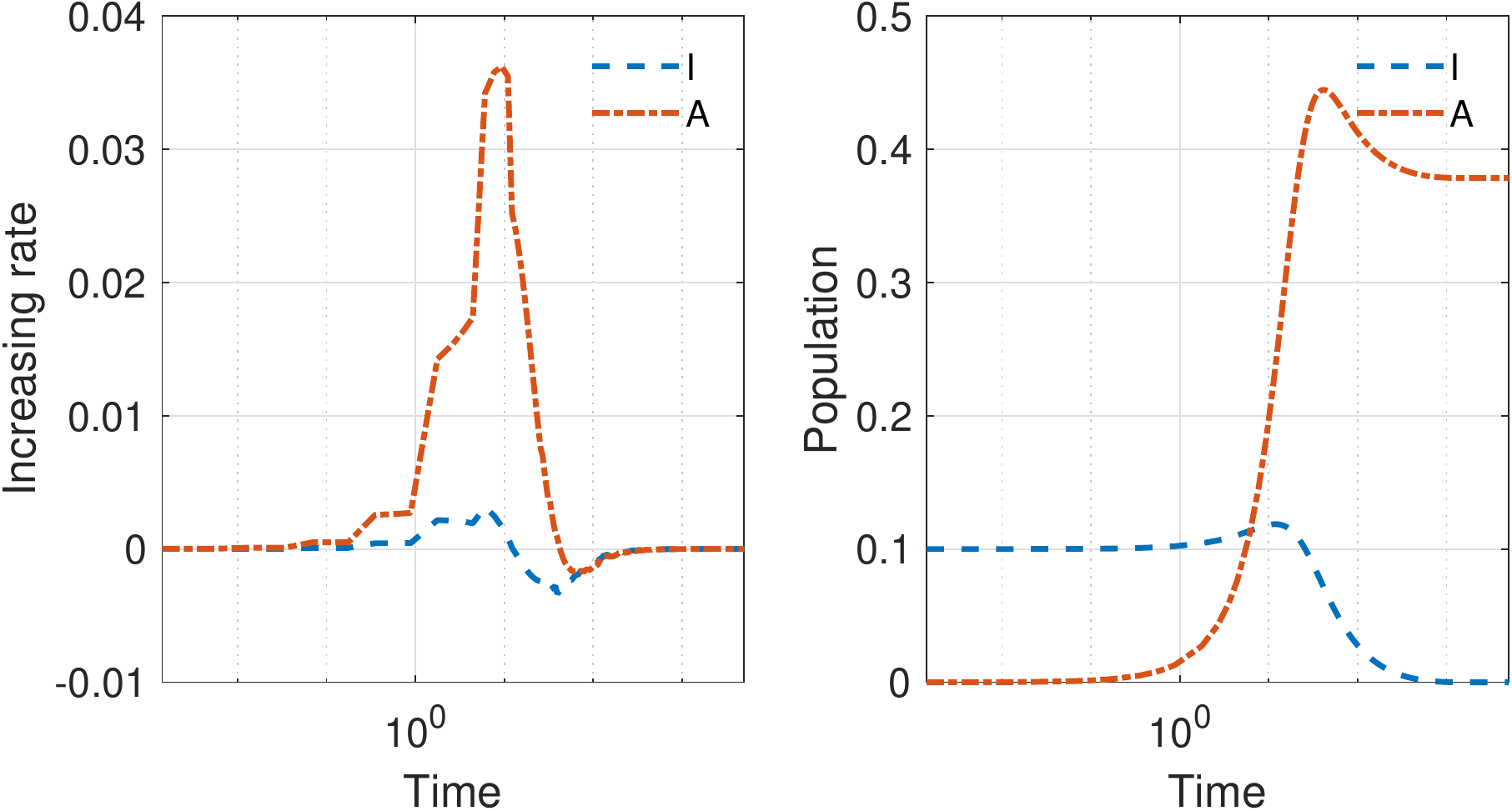}\label{<figure2>}}\\
     \subfloat[Endemic]{\includegraphics[width=0.5\linewidth]{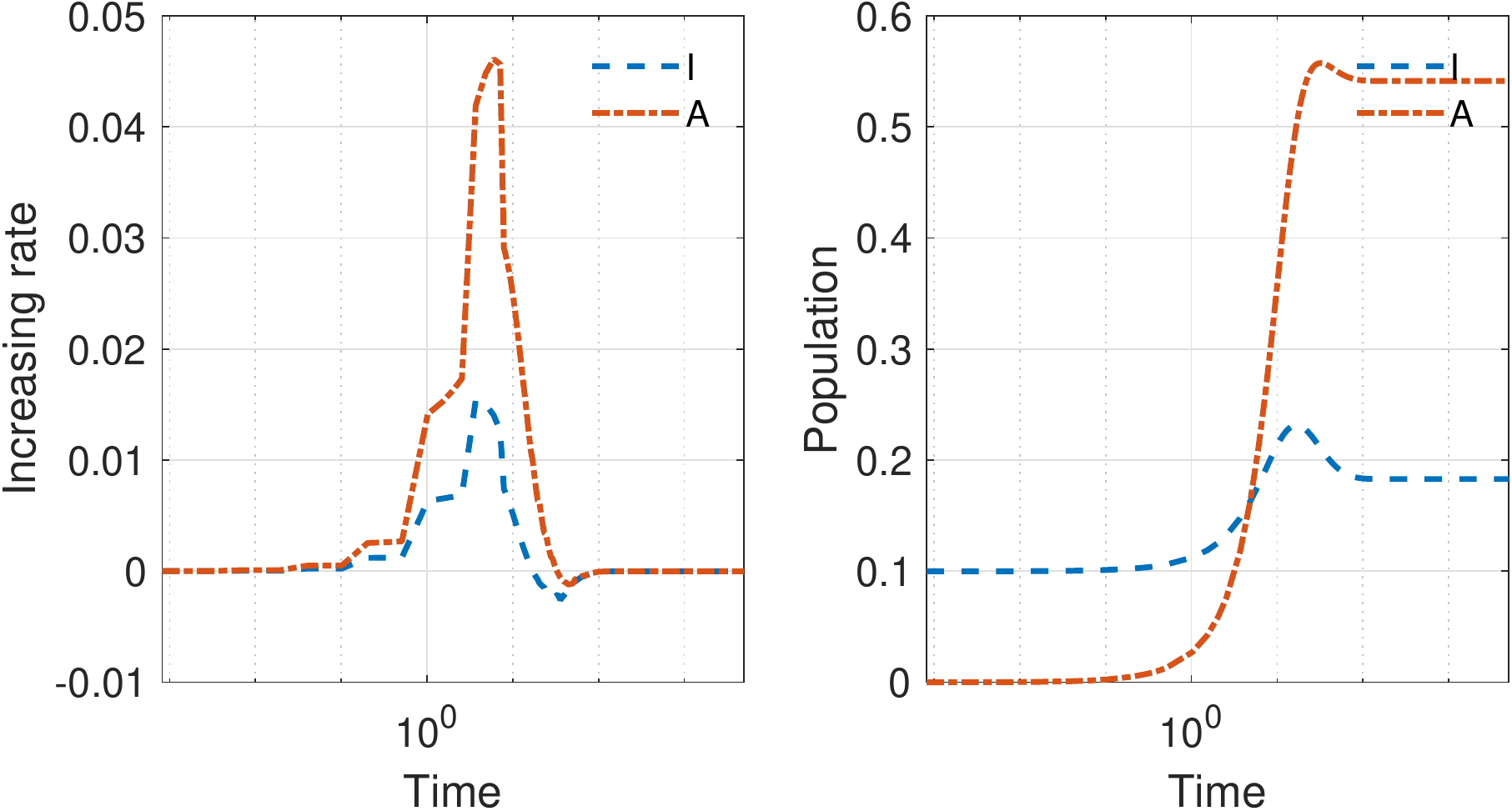}\label{<figure2>}}
    \caption{The rate of change for S and I in three different equilibrium states}
    \label{fig:synchronization}
\end{figure}

Figure~\ref{fig:synchronization} presents the relative growth rates and dynamics of I and A population under different equilibrium states. For both DFE-A and DAFE, the information population will reach its peak shortly after the increase rate of the infectious population reaches its minimum. Note that the synchronizations and disease dynamics during DFE-A and DAFE are different even though the disease gets eliminated in both cases. Under DAFE, the increasing rate of disease is, in general, faster than that of awareness. When the infectious population reaches zero, we observe that the amount of information also starts to drop. This implies that the growth of awareness population is primarily driven by the disease spreading itself, and the strength of the information awareness is not strong enough to persist. On the other hand, under DFE-A, we observe that the growth of awareness population is faster than the infectious population, and the relative growth rate of awareness population is always positive. Moreover, the increase rate also drops with decreasing number of infectious population. Nevertheless, we observe that the growth rates of awareness and the infectious population are positively correlated, and there exists a time lag for the two growth rates to be positively correlated. This time lag is found to be shorter under DAFE, and much longer under DFE-A. Finally, the synchronizations and dynamics when the disease is endemic also differ from both DAFE and DFE-A. In particular, the growth of awareness and the growth of infectious populations are almost perfectly synchronized, but with awareness always spreading faster than the disease. This finding is found to be consistent with the real world observation reported in~\cite{wang2016suppressing}, where the growth rate of patient visit data (corresponding to infectious population) is perfectly synchronized with the trend of Google Flu index (corresponding to awareness population). Based on these findings, we conclude that, when there are many people aware of the risk of the disease, the disease either should be either of minimum risk or it has almost reached its endemic state. 

\begin{figure}[H]
    \centering
    \includegraphics[width=.7\linewidth]{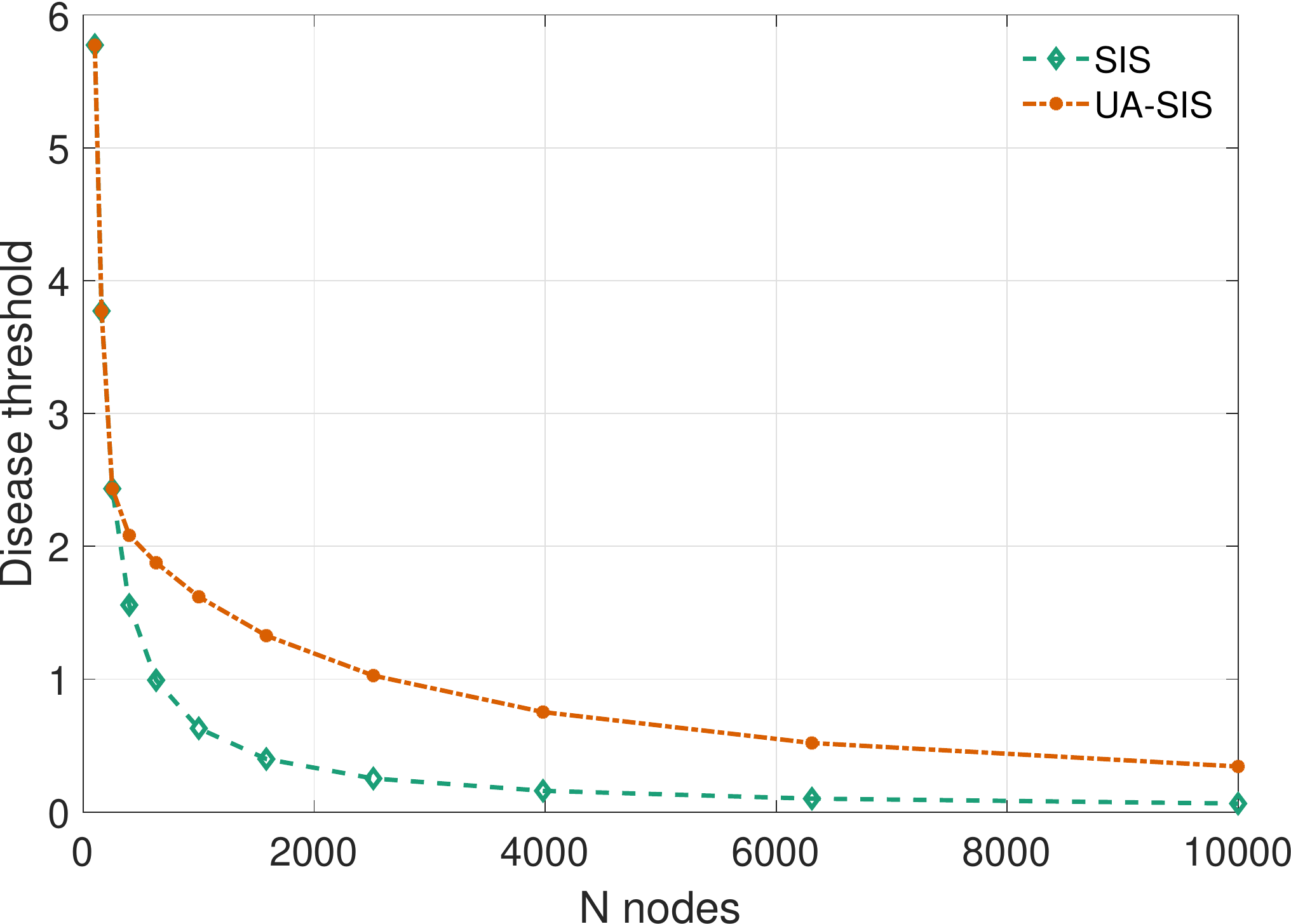}
    \caption{Decay of disease threshold in MCN with increasing number of nodes}
    \label{fig:threshold_scalefree}
\end{figure}
Finally, Figure~\ref{fig:threshold_scalefree} presents how disease threshold changes with the growing size of the network, for both SIS and UA-SIS models. An important and well-known finding for scale-free networks is that there is a lack of disease threshold if the power term of the degree distribution $2\leq\gamma<3$. The reason behind this is that the moment of the variance of degree distribution diverges with the growing size of the network. As discussed earlier, $<k^2>$ also diverges for MCN as the network size grows (equivalent to the increase in the number of travelers). And such a divergence leads to the decay of the disease threshold to 0 as shown in Figure~\ref{fig:threshold_scalefree}, so that the system may become extremely vulnerable to the risk of infectious diseases. Nevertheless, with the help of information dissemination and local awareness through observing other travelers, even a small amount of local information will significantly improve the resilience of the network and significantly delay the decay of the threshold. We observe that the disease threshold under UA-SIS model may be several magnitudes higher than that of the SIS model, especially for very large-sized networks. This observation highlights the importance of incorporating the change of behavior of travelers when we model the dynamics of the infectious diseases and supports the value of the proposed UA-SIS model in better understanding the actual trajectories of an infectious disease. This is a particularly important observation and articulates the effectiveness of local information in preventing target attacks.  

%% file: conclusion.tex
\section{Conclusion}
In this study, the multiplex network model for modeling the co-evolution of information and disease dynamics over the networks is presented. In particular, travelers are assumed to change their behavior based on their observations of the states of their neighbors and by obtaining information from global sources such as news agencies and social media. This percolation of information will have a direct impact on the disease dynamics over the disease network. Meanwhile, the state of the disease spreading also affects the level of information released by global sources and the state and behavior of each individual travelers. The HMF method is used to model the co-evolution of the two dynamics, and obtained three possible stable states. Based on these findings, threshold values for disease and information percolation that may result in one of the three stable states are also discussed and validated by the numerical experiments. 